\documentclass[prl,twocolumn,showpacs,superscriptaddress]{revtex4}
\usepackage{epsfig}
\begin{document}

\title{From non-Brownian Functionals to a Fractional Schr\"odinger Equation}
                                                                                
\author{Lior Turgeman}
\author{Shai Carmi}
\author{Eli Barkai}
\affiliation{Department of Physics, Bar Ilan University, Ramat-Gan
52900 Israel}


\pacs{02.50.Ey,05.40.-a,45.10.Hj}

\begin{abstract}

 We derive
backward and forward
fractional Schr\"odinger type of equations
for the distribution of 
functionals of the path of a particle undergoing anomalous
diffusion. 
Fractional substantial derivatives introduced by
Friedrich and co-workers [PRL {\bf 96}, 230601 (2006)] provide
the correct fractional framework for the problem at hand. 
In the limit of normal diffusion we recover the Feynman-Kac treatment
of Brownian functionals. 
For applications, we calculate the distribution of occupation
times in half space
and show how statistics of anomalous
functionals is related to weak ergodicity breaking. 
\end{abstract}
\maketitle

 Brownian functionals have many applications in physics,
and hence are well investigated \cite{Kac,Majumdar}. With the path of a Brownian
particle $x(t)$ in the time interval $(0,t)$, we define the functional
as $A= \int_0 ^t U[x(t)] {\rm d} t$, where $U(x)$ is some prescribed function.
Since $x(t)$ is a random path, $A$ is a random variable. 
As discussed in a recent review \cite{Majumdar}, 
Brownian functionals model many phenomena, 
such as:
fluctuating interfaces ($U(x)=x^2$), statistics of occupation
times ($U(x)$ is equal to the step function $\theta(x)$), 
Obukhov's model of advection of particles in  turbulent flow 
\cite{Obukhov,Baule1}  
($U(x)=x$ and the velocity field is modeled with Brownian motion),
and finance of stock prices  ($U(x)=\exp(-\beta x)$) \cite{Majumdar},
to name only a few examples. 
Kac used
Feynman's path integral method to obtain the (imaginary time)
Schr\"odinger equation
for the distribution function of $A$ \cite{Kac,Majumdar}. 
 The celebrated Feynman-Kac formula  is 
based on the assumption that
the diffusion is normal Brownian motion. 
However, we know today that in a vast number of applications 
in physics 
the underlying processes exhibit anomalous diffusion 
\cite{Havlin,Bouchaud,Review}.
Hence, the whole machinery of treating Brownian functionals
should be modified for anomalous cases. 

Several authors have recently investigated specific
functionals of anomalous processes. Examples include local
and occupation times ($U(x)= \delta(x)$, $U(x) =\theta(x)$, respectively)
in random walks in the Sinai model
\cite{comtet},  
a fractional Obukhov model  $(U(x)=x)$
\cite{Baule1}, and statistics of occupation times $(U(x)=\theta(x))$
of a particle undergoing sub-diffusive process in half space \cite{JSP}
(see more details below). 
In this manuscript we obtain
 a rather general framework for anomalous functionals,
thus generalizing the Feynman-Kac formalism to an important class 
of stochastic processes. 
We use the continuous time random walk model \cite{Scher},
which is a well established
model 
for anomalous diffusion with many applications in physics
\cite{Bouchaud,Review,Scher}.
Specifically, we will soon derive forward and backward
fractional Schr\"odinger equations, where the
ordinary time derivative is replaced by a fractional substantial derivative \cite{Friedrich},
and the function $U(x)$ plays the role of a potential. 
A few applications are then worked out. 

{\em CTRW model.} We consider a random walk on a one dimensional
lattice with lattice spacing $a$. Jumps are to nearest neighbors
only and with equal probability of jumping left or right. Waiting
times between jump events are independent identically distributed
random variables with a  probability density function (PDF)
$\psi(\tau)$. Thus, the particle waits on lattice point $x_0$ (the starting
point) for time $\tau$ drawn from $\psi(\tau)$, then jumps
with probability $1/2$ to $x_0 + a$ or $x_0 - a$, and then the process
is renewed. The main interest of this Letter is the
case $\psi(\tau) \sim B_\alpha \tau^{ - (1 + \alpha)} / |\Gamma(-\alpha)|$, 
when $0<\alpha<1$. In this case, the waiting time is moment-less 
$(\langle \tau \rangle = \infty)$ and the diffusion becomes
anomalous.  Values of $\alpha$ for  
a large number of systems and models are given in  
\cite{Bouchaud,Review,Scher}. 

{\em CTRW Functionals}.
  Let $G(x,A,t)$ be the joint probability density function of
finding the particle on $(x,A)$ at time $t$. Denote the time the particle
performed the last jump in the sequence as $t-\tau$.  According
to the model the particle is
on $(x,A)$ at time $t$
if it was
on $(x,A- \tau U(x))$ at time $t-\tau$, 
immediately after the last jump was made.
Let $Q_n(x,A,t){\rm d} x {\rm d} A $ be the probability per unit time,
to arrive in $[(x,x+{\rm d} x), (A,A+{\rm d} A)]$ after $n$ jumps.
 Then 
\begin{equation}
G(x,A,t) = \int_0 ^t W\left( \tau \right) \sum_{n=0} ^\infty Q_n \left[x, A- \tau U(x), t-\tau\right] {\rm d} \tau, 
\label{eq01}
\end{equation}
where $ W(\tau)= 1 - \int_0 ^\tau \psi(\tau') {\rm d} \tau'$ 
is the probability
for {\it not} moving in time interval $(t-\tau,t)$. 
 The summation over $n$  is over the random 
number of jumps made.  
The probability of arriving in $(x,A)$ after $n+1$ jumps $Q_{n+1}(x,A,t)$,
is recursively  related to $Q_n(x,A,t)$ 
according to 
\begin{equation}
\begin{array}{l} 
Q_{n+1} (x,A, t) = \int_0 ^t \psi(\tau) {1 \over 2} \left\{ 
Q_n \left[x - a, A - \tau U\left(x-a\right), t-\tau \right] \right. \\
\left. +
Q_n \left[ x+ a , A - \tau U \left( x + a \right) , t- \tau \right] \right\} {\rm d} \tau.
\end{array}
\label{eq02}
\end{equation}
%
%
%
To derive this equation, we notice that to reach point $x$ after $n+1$ steps
 one has
to be after $n$ steps,
 in one of the nearest neighbors ($x-a$ or $x+a$) with probability
$1/2$ for each event. Hence to reach $A$, one has to jump from
$A-\tau U(x\pm a)$, where here
$\tau$, the time interval between jumps, is
randomly distributed with the PDF $\psi(\tau)$. 

 In what follows we consider $U(x) \ge 0$, namely functionals
with positive support, where it is natural to introduce the Laplace
transform $A \to p$ \cite{remark0}. 
It is easy to see that 
\begin{equation}
\int_0 ^\infty Q_n[x, A- U(x) \tau, t-\tau] e^{ - p A} {\rm d} A=
e^{ -p U(x) \tau} Q_n(x, p, t-\tau),
\label{eq03}
\end{equation}
where along this work we use the convention that the variables
in the parenthesis define the space we are working in, and thus
$Q_n(x,p,t)$ is the Laplace $A \to p$ transform of $Q_n(x, A, t)$. 
We now Laplace transform 
Eq. (\ref{eq01}) with respect to time, $t \to s$,
using the  convolution theorem and Eq. (\ref{eq03})
to find
\begin{equation}
G\left(x,p,s\right) = \sum_{n=0} ^\infty { 1  - \hat{\psi} \left[ s + p U(x) \right] \over s + p U(x) } Q_n \left(x,p,s\right), 
\label{eq04}
\end{equation}
where $\hat{\psi}(s) = \int_0 ^\infty \psi(t) \exp( - s t) {\rm d} t$ 
is the Laplace
transform of the waiting time PDF. Fourier  transform,
$x \to k$, of Eq. (\ref{eq04})
yields
\begin{equation}
G(k,p,s) = \sum_{n=0} ^\infty { 1 - \hat{\psi} \left[ s + p U\left( - i {\partial \over \partial k} \right) \right] \over s + p U\left( - i {\partial \over \partial k} \right) }
Q_n (k,p,s ), 
\label{eq05}
\end{equation}
 where the well known Fourier transformation 
$x \to - i {\partial \over \partial k}$ 
was used. 
To complete this part of the derivation we calculate $Q_n (k,p,s )$
using Eq. (\ref{eq02})  
and find the iteration rule
\begin{equation}
Q_{n+1} \left( k , p, s\right) = \cos(k a) \hat{\psi} \left[ s + p U\left(- i {\partial \over \partial k}\right) \right] Q_n (k,p,s).
\label{eq06}
\end{equation}
The initial input for $n=0$ is $Q_0(k,p,s) =1$, since 
$Q_0(x,A,t) = \delta(x)\delta(A) \delta(t)$, namely the particle
is on $x=0$ and $A=0$ at time $t=0$, when the process begins. 
The iteration rule gives
$Q_1 (k,p,s) = \cos(k a) \hat{\psi} (s+ p U(0))$.
 This makes perfect sense since, just after the
first jump we have $A=U(0) \tau$ and $x$ is either on $+a$ or $-a$
(which comes from the inverse Fourier transform of $\cos(ka)$). Notice
that the order of the operators in Eq. (\ref{eq06}) is 
important, since $\cos(ka)$ does not commute with 
$\hat{\psi} \left[ s + p U(- i {\partial \over \partial k} ) \right]$.
 This order of operators is natural, since 
in CTRW we first wait
and then make a jump.
Summing Eq. (\ref{eq05}) over the number of jumps $n$, and using
Eq. (\ref{eq06}), we find 
the formal solution
$$ G(k,p,s) = $$
\begin{equation}
{ 1 - \hat{\psi} \left[ s + pU(- i {\partial \over \partial k} ) \right] \over s + p U \left( - i {\partial \over \partial k} \right)} { 1 \over 1 - \cos(ka) \hat{\psi} \left[ s + p U\left( - i{\partial \over \partial k} \right) \right]}.
\label{eq07}
\end{equation}
When $p=0$, we find the well known Montroll-Weiss
equation \cite{Montroll,Review}
\begin{equation}
G(k,p=0,s) = {1 - \hat{\psi}(s) \over s} {1 \over 1 - \cos(ka) \hat{\psi}(s)},
\label{eq08}
\end{equation}
since 
$G(k,p=0,s)$ 
is the Fourier-Laplace transform of the PDF of finding the particle
on $x$ at time $t$. 

{\em Derivation of forward fractional Schr\"odinger equation.}
As mentioned in the introduction
we assume that the underlying process is
described by power law waiting times with diverging first moment. For that
case we have the small $s$ expansion
\begin{equation} 
\hat{\psi}(s) = 1 - B_{\alpha} s^\alpha + \cdots.
\label{eq09}
\end{equation}
As well known in this case, the diffusion is anomalous with
$\langle x^2 (t) \rangle = 2 K_{\alpha} t^\alpha/\Gamma(1+\alpha)$, 
and 
$K_\alpha= a^2/ 2 B_\alpha$ 
(units $\mbox{mt}^2/\mbox{sec}^\alpha$)
\cite{BarkaiPRE}.
 Substituting Eq. (\ref{eq09}) in 
Eq. (\ref{eq08}) we find the long wavelength $k\to 0$ and long time
$s \to 0$ limit
\begin{equation}
G(k,p,s) \sim \left[ s + p U(- i {\partial \over \partial k} ) \right]^{\alpha -1} {1 \over K_\alpha k^2 +  \left[s + p U( - i {\partial \over \partial k}) \right]^\alpha }.
\label{eq10}
\end{equation}
Rearranging the expression in the last equation we find
\begin{equation}
\begin{array}{l} 
\left[ s + p U(-i {\partial \over \partial k} ) \right]G(k,p,s) - 1 =\\ 
- K_\alpha k^2 \left[ s + p U( - i {\partial \over \partial k} ) \right]^{1-\alpha} G(k,p,s) .
\end{array}
\label{eq11}
\end{equation}
Inverting to the space- time domain $s\to t$ and $k \to x$ 
we find the fractional Schr\"odinger equation
\begin{equation}
{\partial G(x,p,t) \over \partial t} = K_\alpha {\partial^2 \over 
\partial x^2} \  _0 {\cal D} _t ^{1-\alpha} G(x,p,t) - p U(x) G(x,p,t).
\label{eq12}
\end{equation} 
The operator $_0 {\cal D} _t ^{1 - \alpha}$
is a fractional Riemann-Liouville substantial derivative, 
introduced by Friedrich and coworkers  \cite{Friedrich}.
 In Laplace space,  
$_0 {\cal D} _t ^{1 - \alpha} \to (s + p U(x) )^{1-\alpha}$.
When $\alpha=1$, Eq. (\ref{eq12}) is the Schr\"odinger type of equation
for Brownian functionals, originally derived from Feynman-Kac formula 
\cite{Majumdar}.
Also note that if $p=0$, Eq. (\ref{eq12}) reduces to the expected
fractional diffusion equation \cite{Review}. 
Using Eq. (\ref{eq12}),
it is easy to show that
${\partial \langle A \rangle \over \partial t} = \langle U(x) \rangle$, where
$\langle U(x) \rangle= \int U(x) P(x,t) {\rm d} x$, and $P(x,t)$
is the PDF of finding the particle on $x$, i.e.
the solution of the fractional diffusion equation. Of course this  
is the expected behavior since $\partial A / \partial t = U[x(t)]$.

{\em Backward Schr\"odinger equation.}
We now derive a backward equation which turns out to be very useful \cite{remark3}.
Let $G_{x_0}(A,t)$ be the PDF of the functional $A$, when the
process starts on $x_0$. According to the CTRW model, the  particle, 
 after its first
jump at time $\tau$,  goes through either $x_0 +a$ or
$x_0-a$. Alternatively,
 the particle does not move at all during the measurement
time $(0,t)$. Translating this observation to an equation, we have
\begin{widetext}
\begin{equation}
G_{x_0}(A,t) = \int_0 ^t {\rm d} \tau \psi(\tau) {1 \over 2} \left\{
G_{x_0+a}\left[A- \tau U(x_0), t- \tau\right] +
G_{x_0 - a} \left[A- \tau U(x_0), t- \tau\right] \right\}
+ W(t) \delta\left[A - U(x_0) t \right],
\label{eq13aa}
\end{equation} 
\end{widetext}
where $\tau U(x_0)$ is the contribution to $A$ from the pausing time
on $x_0$ in the time interval $(0,\tau)$. The last term on the right hand
side of Eq. (\ref{eq13aa}) describes motionless particles, 
hence the Dirac delta function. 
Using Laplace transform technique similar to that used in
the  derivation of the
forward equation, we find in the continuum limit,
the backward fractional Schr\"odinger equation
\begin{equation}
{\partial G_{x_0} (p,t) \over \partial t} = K_{\alpha}\ _0 {\cal D}_t ^{1-\alpha} {\partial^2 \over \partial (x_0)^2} G_{x_0} (p,t) - p U(x_0) G_{x_0}(p,t).
\label{eq12b}
\end{equation}
Here, the fractional substantial derivative
 $_0 {\cal D}_t ^{1-\alpha}$ is in Laplace $t \to s$ space
$(s + p U(x_0) )^{1-\alpha}$. 
Notice that this operator appears  to the left of the Laplacian 
${\partial^2  / \partial (x_0)^2 }$ in Eq. (\ref{eq12b}), in contrast
to the forward equation (\ref{eq12}).
Since in Eq. (\ref{eq12b})  the operators  depend on 
$x_0$ and not on $x$, Eq. (\ref{eq12b})  is called a backward equation. 
Eqs. (\ref{eq12},\ref{eq12b}) are the main equations of this manuscript
since they provide a general framework for treating functionals
of anomalous processes. When $\alpha=1$ both equations  reduce to
the usual Schro\"dinger
equations found in the Feynman-Kac treatment of Brownian functionals
\cite{Majumdar}.

{\em  First Illustration: Lamperti's law for
distribution of occupation times}. We consider the occupation
time in half space: $A= \int_0 ^t \theta[x(\tau)] {\rm d} \tau$,
usually denoted with $t^{+}$.
This distribution was first computed by Lamperti using probabilistic
methods \cite{Lamperti}  (see also \cite{JSP}).
Substituting $U(x_0)= \theta(x_0)$ in Eq.  (\ref{eq12b}), 
we solve the backward equation separately for $x_0>0$ and
$x_0<0$, demanding the continuity of the solution and
its derivative. In addition we have 
$\lim_{x_0 \to \infty} G_{x_0} (p,s) = (s + p)^{-1}$ 
since if $x_0 \to \infty$,
the particle is always in $x>0$ and thus $A=t$. Similarly 
$\lim_{x_0 \to - \infty} G_{x_0} (p,s) = s^{-1}$ since then
$A=0$. 
At least in
Laplace space, the solution of the fractional equation 
is easily obtained.
For initial condition $x_0=0$, we find, after inversion of the solution
to the time domain, the PDF of the scaled functional $p^{+}=t^{+}/t$ i.e.
the fraction of time spent in half of the space
\begin{equation}
\begin{array}{l}
f(p^{+} ) =  \\
{ \sin \pi {\alpha \over 2} \over \pi} { \left( p^{+}\right)^{\alpha/2 -1} \left( 1 - p^{+} \right)^{\alpha/2 -1} \over \left( 1 - p^{+} \right)^\alpha + \left( p^{+} \right)^\alpha + 2 \left( 1 - p^{+} \right)^{\alpha/2} \left( p^{+} \right) ^{\alpha/2} \cos {\pi \alpha \over 2} }.
\end{array}
\label{eqLamp}
\end{equation}
Naively, one expects that the particle spends half of
the time in $x>0$. 
 In contrast, the Lamperti  PDF Eq. (\ref{eqLamp}) has two peaks at
$p^{+}=1$ and $p^{+} = 0$, while its minimum on $p^{+} =1/2$
coincides with its average
$\langle p^{+}\rangle =1/2$. 
In the
limit $\alpha \to 0$ we get two delta functions on $p^{+} = 1$
and $p^{+} = 0$, indicating that the particle is localized
in either $x>0$ or $x<0$ for the whole observation time.
For $\alpha \to 1$ we recover the well known arcsine law of P. L\'evy
\cite{Majumdar}.

{\em Second Illustration: Weak Ergodicity breaking.} 
The technique used so far is valid for free sub-diffusion.
Our approach can be easily extended to study functionals in
an external binding field, to give new insights on the problem
of ergodicity. We consider CTRW in the harmonic potential 
$V(x) = m \omega^2 x^2 /2$ 
as modeled by the fractional Fokker-Planck equation \cite{Review,MBK,BarkaiPRE}.
(see \cite{Bray} for normal diffusion). 
Algorithms that generate the stochastic continuous trajectories
were recently proposed
\cite{Weron,Friedrich1}. 
Consider the time average
$\overline{x} = \int_0 ^t x(\tau) {\rm d} \tau/t$, hence
$A=\int_0 ^t x(\tau) {\rm d} \tau$ and $U(x) =x$
\cite{remark0}. 
If the process is ergodic, then, due to the symmetry of the harmonic field, 
the time average $\overline{x} = \int_0 ^t x(\tau){\rm d }\tau/ t$, in the long
time limit, is statistically equal to zero. For anomalous
sub-diffusion  $(\alpha<1)$ it is well known that
ergodicity is broken \cite{WEB1,WEB2}. With our tool box for
anomalous functionals, we now treat the problem of fluctuations of
the time average $\overline{x}$.

\begin{figure}
\begin{center}
\epsfig{figure=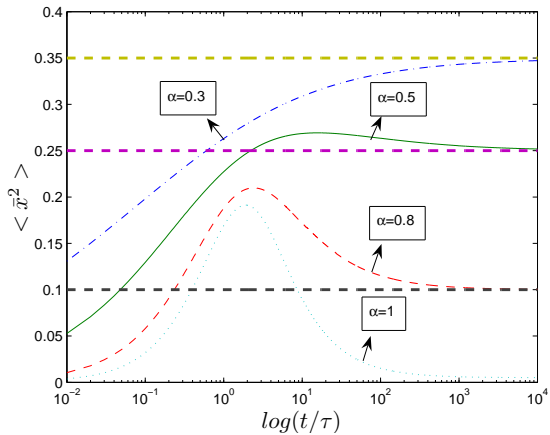, width=0.45\textwidth}
\end{center}
\caption{ Fluctuations of the time average of $x(t)$, $\langle \overline{x}^2 
\rangle$ versus $t/\tau$ 
for a particle in a binding Harmonic field, do not vanish
when $t \to \infty$ thus ergodicity is broken. The exception is the
case of normal diffusion $\alpha=1$, which exhibits ergodic behavior. 
The dashed straight lines are the limit $\lim_{t \to \infty} \langle \overline{x}^2 \rangle=(1-\alpha) \langle x^2 \rangle_{th}$.
We use $\langle x^2 \rangle_{th} = 1/2$, $x_0=0$ and hence 
$\langle \overline{x}^2 (t=0) \rangle = (x_0)^2 =0$ and according
to Eq. (\protect{\ref{eq14w}})  for $\alpha<1/3$ $(\alpha>1/3)$
solution approaches the asymptotic limit from below (above).
}
\label{fig1}
\end{figure}

%
%
We use the forward Eq. (\ref{eq12})  for $A=\int_0 ^t x(\tau) {\rm d} \tau$
with the modification
that  the Laplacian of free diffusion, 
is replaced by the Fokker-Planck operator 
$L_{{\rm fp}} = K_{\alpha} ( {\partial \over \partial x} m \omega^2 x / k_b T + {\partial^2 \over \partial x^2} )$,
which will be further justified in a longer publication 
and $T$ is the temperature. 
Thus, the modified forward equation of motion is 
\begin{equation}
{\partial G(x,A,t) \over \partial t} 
=
L_{{\rm fp}}\ _0 {\cal D}_t ^{1-\alpha} G(x,A,t)
 - x {\partial \over \partial A}.  
G(x,A,t)
\label{eq14}
\end{equation}
Here in Laplace $t\to s$ space we have $_0 {\cal D}_t ^{1-\alpha} \to
(s + x {\partial \over \partial A})^{1-\alpha}$. 
The same equation, after renaming of variables, describes
a weakly damped kinetic model of super-diffusive L\'evy walks 
\cite{Friedrich,remark2}.
We investigate the fluctuations
of the time average, namely, 
$\langle \overline{x}^2 \rangle = \langle A^2 \rangle/t^2$. 
Using Eq. (\ref{eq14}), we derive equations of motion for the
 low order moments
$\langle x \rangle$, $\langle A \rangle$,
 $\langle x A \rangle$, $\langle A^2 \rangle$ and $\langle x^2 \rangle$.
 These equations are closed in the sense that they are not coupled
to any higher order moments, due to the harmonic potential 
under investigation and the choice of the functional.  Using the relaxation time
 $\tau^\alpha \equiv k_b T/(K_\alpha m \omega^2)$, we find in
 Laplace space
\begin{widetext} 
\begin{equation}
\langle A^2 \rangle = {2 \over s^3} { \left[ \left(1 - \alpha \right) + \left( s \tau\right)^\alpha \right] \over 1 + \left( s \tau \right)^\alpha  } { \left[ 2 \langle x^2 \rangle_{th} + \left( s \tau \right)^\alpha \left( x_0 \right)^2 \right] \over
2 + \left( s \tau\right)^\alpha },
\end{equation} 
where
$\langle x^2 \rangle_{th} =
{k_b T/ m \omega^2}$ and $x_0$ is the initial condition. 
Inverting to the time domain, using 
$\langle \overline{x}^2 \rangle = \langle A^2 \rangle /t^2$,
we find
\begin{equation}
\langle \overline{x}^2 \rangle = \left( 1 - \alpha \right) \langle x^2 \rangle_{th} + 2 \alpha \left( 2 \langle x^2 \rangle_{th} - (x_0)^2 \right)
E_{\alpha,3} \left[ - \left( t/\tau\right)^\alpha \right] +
2 \left( 1 + \alpha\right) \left( (x_0)^2 - \langle x^2 \rangle_{th} \right) E_{\alpha,3} \left[ - 2 \left( t /\tau \right)^\alpha \right],
\label{eq14x}
\end{equation}
where $E_{\alpha,3}(z) = \sum_{n=0} ^\infty z^n / \Gamma(\alpha n  + 3 )$
is the Mittag-Leffler function 
\cite{Podlubny}. To derive Eq. (\ref{eq14x}) we used
$\int_0 ^\infty \exp( - s t) t^2 E_{\alpha, 3} [ -c t^\alpha] {\rm d} t = s^{\alpha - 3 }/ (s^\alpha + c )$ \cite{Podlubny}.    
For long times we have
\begin{equation}
\langle \overline{x}^2  \rangle \sim \left( 1 - \alpha \right) \langle x^2 \rangle_{th} + \left[ \left( 3 \alpha - 1 \right) \langle x^2 \rangle_{th} + \left( 1 - \alpha\right) (x_0)^2 \right] \Gamma(3 -\alpha)^{-1}\left( { \tau \over t} \right)^\alpha+ O\left(\left(\frac{\tau}{t}\right)^{2\alpha}\right),
\label{eq14w}
\end{equation}
\end{widetext}
hence 
$\lim_{t \to \infty} \langle \overline{x}^2 \rangle=\langle x^2 \rangle_{th} (1 - \alpha)$ in agreement with \cite{WEB2}.
Only when $\alpha=1$ we have ergodic behavior
$\lim_{t \to \infty} \langle \overline{x}^2 \rangle=0$.
Eqs. (\ref{eq14x},\ref{eq14w}) give, for the first time, the time dependence of
$\langle \overline{x}^2 \rangle$ which exhibits a 
$t^{-\alpha}$ power law convergence 
to 
the long time limit. The convergence to the asymptotic result is
extremely slow when $\alpha\to 0$ as might be expected. 
Notice that the
initial condition $x_0$ enters in the correction term to the long time
limit,
and in this sense initial conditions decay slowly.  
For short times we have 
$\langle \overline{x} ^2 \rangle \simeq (x_0)^2 + [4 (\langle x^2 \rangle_{\rm th} - (x_0)^2)- 2 \alpha (x_0)^2] (t/\tau)^\alpha/\Gamma(3 + \alpha)$. 
We demonstrate these behaviors in 
Fig. \ref{fig1} where $\langle \overline{x}^2 \rangle$ versus
$t$ is plotted for various values of $\alpha$.  

%
 To conclude we have found both  forward and backward fractional
 Schr\"odinger equations
describing distributions of
 functionals of widely observed sub-diffusive processes. 
When a binding force is included, the analysis of
these functionals leads to a new kinetic
approach to weak ergodicity breaking, in contrast to the equilibrium
framework provided in \cite{WEB1,WEB2}.    
While previous work considered
specific  functionals, our work 
provides a very general toolbox for this new field.

{\bf Acknowledgment} This work was supported by the Israel Science Foundation.
S.C. is supported by the Adams Fellowship Program of the Israel Academy of
Sciences and Humanities.


\begin{thebibliography}{99}

\bibitem{Kac} M. Kac, {\em Trans. Am. Math. Soc.} {\bf 65} 1 (1949).
M. Kac, Proceedings of the Second Berkeley Symposium on Mathematical
Statistics and Probability, p. 189 (1951). 

\bibitem{Majumdar} S. N. Majumdar, {\em Current Science} {\bf 89} 2076 (2005).

\bibitem{Obukhov} A. M. Obukhov, {\em Adv. in Geophysics} {\bf 6} 113 (1959). 

\bibitem{Baule1} A. Baule,  and R. Friedrich, {\em Physics Letters A} {\bf 350} 167 (2006). 

\bibitem{Havlin} S. Havlin, D. ben-Avraham, {\em Adv. Phys.} {\bf 36}, 695 (1987).


\bibitem{Bouchaud} J. P. Bouchaud, and A. Georges, {\em Phys. Rep.}
{\bf 195}, 127 (1990). 

\bibitem{Review} R. Metzler, J. Klafter, {\em Phys. Rep.} {\bf 339}, 1 (2000).

\bibitem{comtet} S. N. Majumdar, A. Comtet, {\em Phys. Rev. Lett.} {\bf 89}
060601 (2002). 

\bibitem{JSP}
  E. Barkai
{\em J. of Statistical Physics} {\bf 123} 883 (2006).

\bibitem{Scher} H. Scher, and E. Montroll {\em Phys. Rev. B} {\bf 12}, 
2455 (1975).  

\bibitem{Friedrich} R. Friedrich, F. Jenko, A. Baule, S. Eule {\em Phys. Rev. Lett.} {\bf 96} 230601 (2006). 

\bibitem{remark0} 
For Brownian
functionals 
Majumdar \cite{Majumdar}
 considers $U(x)\ge 0$. 
The case where $U(x)$ has negative support can be  
easily treated with
Fourier transform as we will discuss in a future publication. 

\bibitem{Montroll} E. W. Montroll and G. H. Weiss, {\em J. Math. Phys.} {\bf 6}, 167 (1965).


\bibitem{BarkaiPRE} 
E. Barkai, R. Metzler and J. Klafter,
{\em Phys. Rev. E} {\bf 61} 132 (2000).

\bibitem{remark3} To find the PDF of the functional $A$, the solution
of the forward Eq.
(\ref{eq12}) $G(x,p,t)$ must be: i) Laplace inverted $p \to A$ and ii)
integrated over all $x$. To overcome the second complication (get rid
of $x$) the backward equation is useful.  

\bibitem{Lamperti} J. Lamperti,
{\em Trans. Am. Math. Soc.} {\bf 88}, 380 (1958).

\bibitem{MBK} R. Metzler, E. Barkai, and J. Klafter 
{\em Phys. Rev. Lett.} {\bf 82}, 3563 (1999).

\bibitem{Bray} S. N. Majumdar, and A. J. Bray {\em Phys. Rev. E} {\bf 65} 051112 (2002). 

\bibitem{Weron} 
M. Magdziarz, A. Weron, and
K. Weron
{\em Phys. Rev. E} {\bf  75}, 016708 (2007). 


\bibitem{Friedrich1}  
D. Kleinhans and R. Friedrich
{\em Phys. Rev. E} {\bf 76}, 061102 (2007).


\bibitem{WEB1} G. Bel, E. Barkai
{\em Phys. Rev. Lett.} {\bf 94} 240602 (2005).

\bibitem{WEB2}
A. Rebenshtok, E. Barkai, 
{\em Phys. Rev. Lett.} 
{\bf 99}, 210601 (2007). 
{\em ibid} 
{\em Journal of Statistical Mechanics}  {\bf 133} 565 (2008).

\bibitem{remark2} Sub-diffusive motion in a harmonic potential
is sometimes called the fractional
Ornstein-Uhlenbeck process \cite{MBK}. As we show in the
text, it yields the scaling $\langle A^2 \rangle\sim t^2$ 
which leads to ergodicity breaking. 
For the kinetic model of ballistic L\'evy walks in \cite{Friedrich},
the velocity undergoes a fractional
Ornstein-Uhlenbeck process and there $\langle x^2 \rangle\sim t^2$. 

\bibitem{Podlubny} I. Podlubny {\em Fractional Differential Equations} 
(Academic Press, New York, 1999). 

\end{thebibliography}
\end{document}